\documentclass[aps,preprint,nopacs,superscriptaddress,12pt]{revtex4-2}
\usepackage{physics}
\usepackage{gensymb}
\usepackage{xspace}
\usepackage[separate-uncertainty=true,multi-part-units=single]{siunitx}
\usepackage{graphicx}
\usepackage{setspace}

\newcommand{\figtitle}[1]{\textbf{#1}\xspace}
 
\begin{document}
\setstretch{1}
\title{Observation of the dual quantum spin Hall insulator by density-tuned correlations in a van der Waals monolayer}

\author{Jian Tang}
\thanks{These authors contributed equally.}
\affiliation{Department of Physics, Boston College, Chestnut Hill, MA, USA}
\author{Thomas Siyuan Ding}
\thanks{These authors contributed equally.}
\affiliation{Department of Physics, Boston College, Chestnut Hill, MA, USA}
\author{Hongyu Chen}
\thanks{These authors contributed equally.}
\affiliation{Division of Physics and Applied Physics, School of Physical and Mathematical Sciences, Nanyang Technological University, Singapore}
\author{Anyuan Gao}
\affiliation{Department of Chemistry and Chemical Biology, Harvard University, Cambridge, MA, USA}
\author{Tiema Qian}
\affiliation{Department of Physics and Astronomy and California NanoSystems Institute, University of California Los Angeles, Los Angeles, CA, USA}
\author{Zumeng Huang}
\affiliation{Department of Physics, Boston College, Chestnut Hill, MA, USA}
\author{Zhe Sun}
\affiliation{Department of Physics, Boston College, Chestnut Hill, MA, USA}
\affiliation{Department of Chemistry and Chemical Biology, Harvard University, Cambridge, MA, USA}
\author{Xin Han}
\affiliation{Beijing National Laboratory for Condensed Matter Physics and Institute of Physics, Chinese Academy of Sciences, Beijing, China}
\author{Alex Strasser}
\affiliation{Department of Materials Science and Engineering, Texas A$\&$M University, College Station, TX, USA}
\author{Jiangxu Li}
\affiliation{Department of Physics and Astronomy, University of Tennessee, Knoxville, TN, USA}
\affiliation{Min H. Kao Department of Electrical Engineering and Computer Science, University of Tennessee, Knoxville, Tennessee, USA}
\author{Michael Geiwitz}
\affiliation{Department of Physics, Boston College, Chestnut Hill, MA, USA}
\author{Mohamed Shehabeldin}
\affiliation{Department of Physics, Boston College, Chestnut Hill, MA, USA}
\author{Vsevolod Belosevich}
\affiliation{Department of Physics, Boston College, Chestnut Hill, MA, USA}
\author{Zihan Wang}
\affiliation{Department of Physics, Boston College, Chestnut Hill, MA, USA}
\author{Yiping Wang}
\affiliation{Department of Physics, Boston College, Chestnut Hill, MA, USA}
\author{Kenji Watanabe}
\affiliation{Research Center for Electronic and Optical Materials, National Institute for Materials Science, Tsukuba, Japan}
\author{Takashi Taniguchi}
\affiliation{Research Center for Materials Nanoarchitectonics, National Institute for Materials Science,  Tsukuba, Japan}
\author{David C. Bell}
\affiliation{Harvard John A. Paulson School of Engineering and Applied Sciences and The Center for Nanoscale system, Harvard University, Cambridge, MA, USA}
\author{Ziqiang Wang}
\affiliation{Department of Physics, Boston College, Chestnut Hill, MA, USA}
\author{Liang Fu}
\affiliation{Department of Physics, Massachusetts Institute of Technology, Cambridge, MA, USA}
\author{Yang Zhang}
\affiliation{Department of Physics and Astronomy, University of Tennessee, Knoxville, TN, USA}
\affiliation{Min H. Kao Department of Electrical Engineering and Computer Science, University of Tennessee, Knoxville, Tennessee, USA}
\author{Xiaofeng Qian}
\affiliation{Department of Materials Science and Engineering, Texas A$\&$M University, College Station, TX, USA}
\author{Kenneth S. Burch}
\affiliation{Department of Physics, Boston College, Chestnut Hill, MA, USA}
\author{Youguo Shi}
\affiliation{Beijing National Laboratory for Condensed Matter Physics and Institute of Physics, Chinese Academy of Sciences, Beijing, China}
\author{Ni Ni}
\affiliation{Department of Physics and Astronomy and California NanoSystems Institute, University of California Los Angeles, Los Angeles, CA, USA}
\author{Guoqing Chang}
\email{Corresponding emails: maqa@bc.edu, guoqing.chang@ntu.edu.sg\\
Submitted version}
\affiliation{Division of Physics and Applied Physics, School of Physical and Mathematical Sciences, Nanyang Technological University, Singapore}
\author{Su-Yang Xu}
\affiliation{Department of Chemistry and Chemical Biology, Harvard University, Cambridge, MA, USA}
\author{Qiong Ma}
\email{Corresponding emails: maqa@bc.edu, guoqing.chang@ntu.edu.sg\\
Submitted version}
\affiliation{Department of Physics, Boston College, Chestnut Hill, MA, USA}
\affiliation{CIFAR Azrieli Global Scholars program, CIFAR, Toronto, Canada}

\maketitle
\clearpage

\subsection*{Abstract}

\textbf{The convergence of topology and correlations represents a highly coveted realm in the pursuit of novel quantum states of matter~\cite{wen2019choreographed, tokura2022quantum}. Introducing electron correlations to a quantum spin Hall (QSH) insulator can lead to the emergence of a fractional topological insulator and other exotic time-reversal-symmetric topological order~\cite{levin2009fractional, maciejko2010fractional, santos2011time, goerbig2012fractional,li2014fractional, wang2016time, barkeshli2019symmetry,wu2023time}, not possible in quantum Hall and Chern insulator systems. However, the QSH insulator with quantized edge conductance remains rare, let alone that with significant correlations. In this work, we report a novel \textit{dual} QSH insulator within the intrinsic monolayer crystal of TaIrTe$_4$, arising from the interplay of its single-particle topology and density-tuned electron correlations. At charge neutrality, monolayer TaIrTe$_4$ demonstrates the QSH insulator that aligns with single-particle band structure calculations, manifesting enhanced nonlocal transport and quantized helical edge conductance. Interestingly, upon introducing electrons from charge neutrality, TaIrTe$_4$ only shows metallic behavior in a small range of charge densities but quickly goes into a new insulating state, entirely unexpected based on TaIrTe$_4$'s single-particle band structure. This insulating state could arise from a strong electronic instability near the van Hove singularities (VHS), likely leading to a charge density wave (CDW). Remarkably, within this correlated insulating gap, we observe a resurgence of the QSH state, marked by the revival of nonlocal transport and quantized helical edge conduction. Our observation of helical edge conduction in a CDW gap could bridge spin physics and charge orders. The discovery of a \textit{dual} QSH insulator introduces a new method for creating topological flat minibands via CDW superlattices, which offer a promising platform for exploring time-reversal-symmetric fractional phases and electromagnetism~\cite{levin2009fractional, park2010dirac, maciejko2010fractional, park2010dirac, santos2011time, wang2022fractional}.} 

\vspace{-4 mm}
\subsection*{Introduction}

Topology and correlations are two central themes of current condensed matter research~\cite{hasan2010colloquium, tokura2022quantum}. Their interplay gives rise to correlated topological phases~\cite{wen2019choreographed}, including topological order, which could lead to phenomena such as topological fractionalization, long-range entanglement, and non-Abelian anyons. Recent theory developments have highlighted materials exhibiting both single-particle topological bands and strong correlations as exciting candidates to explore in the search for correlated topological phases~\cite{sheng2011fractional, neupert2011fractional, tang2011high, regnault2011fractional}. An intriguing but rarely explored scenario is ``QSH + correlations", i.e., to introduce correlations to the $\mathcal{Z}_2$ band of a QSH insulator~\cite{hasan2010colloquium}. The time-reversal symmetry present for the $\mathcal{Z}_2$ band can enable time-reversal-symmetric topological order such as the fractional QSH and the helical quantum spin liquid~\cite{levin2009fractional, maciejko2010fractional, park2010dirac, santos2011time,goerbig2012fractional,li2014fractional, wang2016time, barkeshli2019symmetry,wu2023time}, unattainable in quantum Hall~\cite{stormer1999fractional} and Chern insulators~\cite{xie2021fractional, cai2023signatures, zeng2023thermodynamic, park2023observation, xu2023observation}.\\ 

Despite the fundamental interest in this ``QSH + correlations" scenario, it is rarely investigated due to the scarcity of suitable material platforms. First, while the QSH state has been explored in many materials~\cite{bernevig2006quantum, konig2007quantum,yang2012spatial, xu2013large, drozdov2014one, qian2014quantum, du2015robust,zhu2015epitaxial, li2016experimental, fei2017edge, tang2017quantum, dong2019observation, shumiya2022evidence,wang2023excitonic}, achieving the QSH effect with quantized helical edge conduction has been extremely rare~\cite{konig2007quantum, du2015robust, wu2018observation, zhao2024realization}. Second, it is even rarer to find a QSH insulator exhibiting strong correlations, as most QSH materials have dispersive bands. One effective way to induce strong correlations involves achieving a large density of states via the van Hove singularity (VHS) (Fig.~\ref{Fig1}\textbf{a-b}), a strategy that has resurged in recent studies of graphene and kagome materials~\cite{wu2021chern, zhou2022isospin, yin2022topological, teng2022discovery}. In this work, we report a new electronic phenomenon that arises from the interplay of QSH topology and VHS-induced correlations, the \textit{dual} QSH insulator, in a vdW monolayer crystal TaIrTe$_4$. Specifically, we detect two distinct charge gaps: one at charge neutrality and another near the DFT-calculated VHS positions (Fig.~\ref{Fig1}\textbf{c}). Each gap exhibits QSH helical edge conduction, named as QSH-I and QSH-II, respectively. Our systematic study shows that while QSH-I represents a single-particle QSH effect, QSH-II is a correlated QSH, likely arising from a topological CDW that unusually links spin transport with charge order. Moreover, this \textit{dual} QSH reveals a flat QSH band, which holds promise for exploring time-reversal invariant fractional phases.

\vspace{-4 mm}
\subsection*{Lattice and electronic structures of monolayer TaIrTe$_4$}

\begin{figure*}
\includegraphics[width=4.5in]{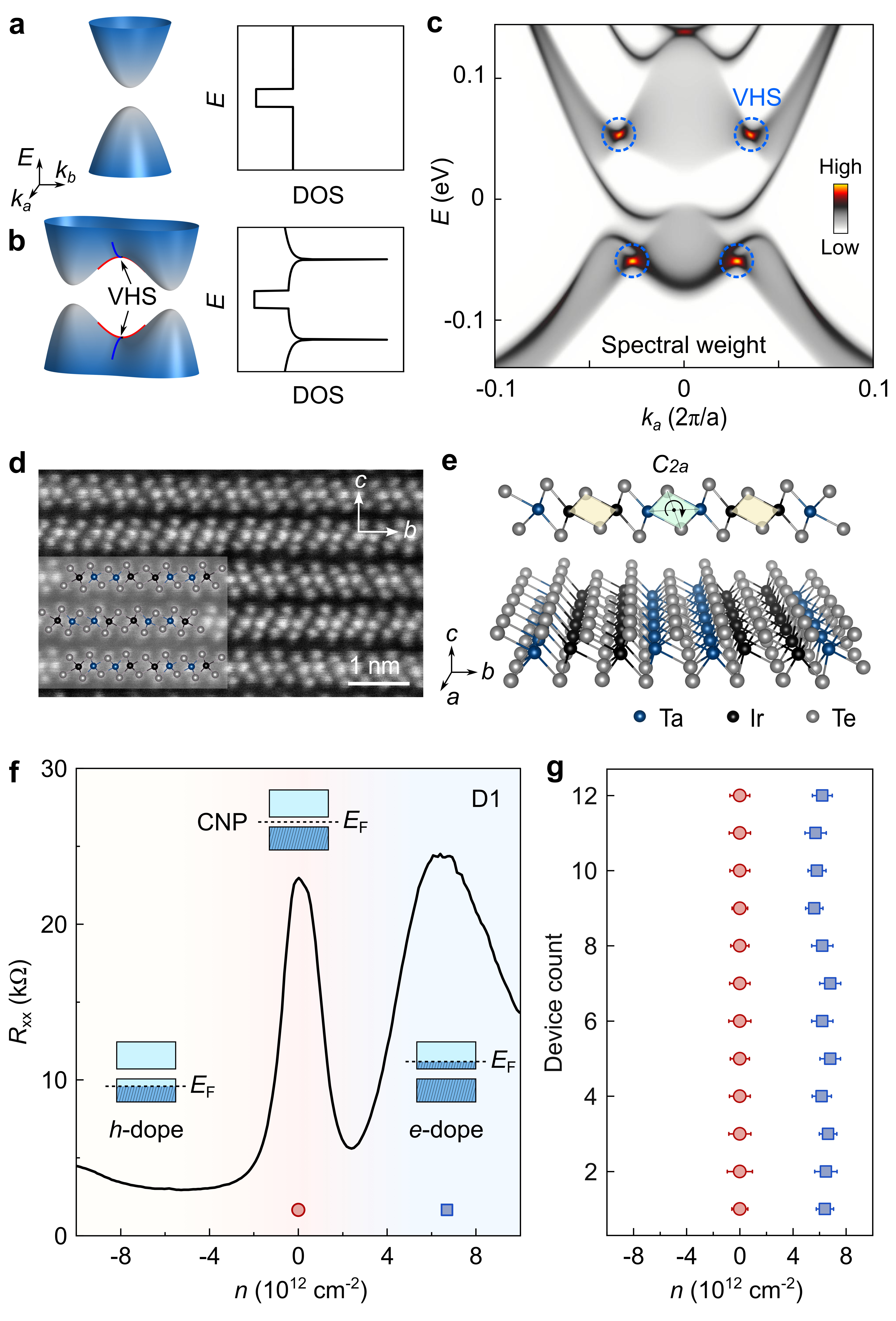}
\caption{{\figtitle{Electronic band structure, lattice structure, and basic electrical characterization of monolayer TaIrTe$_4$.} 
\textbf{a,} Sketch of a simple parabolic band structure featuring constant density of states (DOS) beyond the band gap. 
\textbf{b,} Sketch of two parabolic bands merging to form van Hove singularity (VHS), leading to divergent DOS. 
\textbf{c,} First-principles calculated projected spectral weight along the $k_\mathrm{a}$ direction showing the edge state dispersion and bulk VHS. 
\textbf{d,} Cross-sectional TEM image of the TaIrTe$_4$ crystal. The inset illustrates the overlap with the atomic configuration of TaIrTe$_4$. 
\textbf{e,} Lattice structure and symmetry of monolayer TaIrTe$_4$. 
\textbf{f,} The typical four probe resistance $R_\mathrm{xx}$ versus carrier density $n$ at $T$ = 1.7 K (device D1). The light yellow (blue) shaded region indicates the point at which resistance begins to increase in the hole (electron)-doped regions, respectively. 
\textbf{g,} A summary of the resistance peak positions across a dozen of devices. The error bars are determined based on the point at which the resistance decreases to $\SI{95}{\percent}$ of the peak value.}}\label{Fig1}
\end{figure*}

Bulk TaIrTe$_4$ is a layered (Fig.~\ref{Fig1}\textbf{d}) Weyl semimetal that has been investigated in both theoretical and experimental studies~\cite{liu2018raman, ma2019nonlinear, belopolski2017signatures, cai2019observation, dong2019observation, kumar2021room}. In contrast, monolayer TaIrTe$_4$ has not yet been experimentally explored, due to its severe air-sensitivity and the challenges associated with exfoliation. By systematically optimizing the fabrication conditions (Methods), we have successfully accessed the intrinsic electronic properties of monolayer TaIrTe$_4$ for the first time (see Extended Data Fig. 1 and Supplementary Information (SI) Section 1). \\

Structurally, monolayer TaIrTe$_4$ features 1D Ta and Ir chains along the crystalline $\hat{a}$ axis (Fig.~\ref{Fig1}\textbf{e}). Within a single unit cell, there are two inequivalent Ta (Ir) chains, which are related by a $180^{\circ}$ rotation about the $\hat{a}$ axis ($C_\mathrm{2a}$) (Fig.~\ref{Fig1}\textbf{e}). Observing the $b$-$c$ plane, the monolayer is made of alternating parallelograms - one parallelogram is formed by two Ta and two Te while the other is formed by two Ir and two Te. In terms of electronic properties, theoretical predictions suggest that monolayer TaIrTe$_4$ is a QSH insulator with a substantial band gap~\cite{liu2017van, guo2020quantum}. Notably, our calculations also reveal prominent VHS close to the single-particle QSH gap, as indicated by the bright intensity spots inside the projected bulk bands in Fig.~\ref{Fig1}\textbf{c}. We will revisit this lattice and electronic structures, as it proves crucial for understanding our observations.

\vspace{-5 mm}
\subsection*{Unconventional basic transport characteristics of monolayer TaIrTe$_4$}

We have fabricated high-quality dual-gated (top gate voltage, $V_{\mathrm{tg}}$ and bottom gate voltage, $V_{\mathrm{bg}}$) monolayer TaIrTe$_4$ devices. We observed highly consistent behaviors across over a dozen devices. Transport characterizations conducted on 19 devices indicate mobility ranging from approximately 10 to 400 cm$^2$/(V$\cdot$s) (SI 1.4). Figure~\ref{Fig1}\textbf{f} shows the four-probe resistance $R_\mathrm{xx}$ as a function of the carrier density $n$ (see electric field $E$ dependence and discussions in SI 4.2 and 5.2). Notably, even for this basic property, monolayer TaIrTe$_4$ already distinctly differs from known QSH systems including monolayer WTe$_2$~\cite{fei2017edge, wu2018observation}. Specifically, in addition to the resistance peak at charge neutrality which is expected based on the single-particle band structure, we observe a second resistance peak in the electron-doped regime at $\sim6.5\times10^{12}$ cm$^{-2}$ (Fig.~\ref{Fig1}\textbf{f-g}). Below, we systematically study both resistive peaks.

\vspace{-4 mm}
\subsection*{Observation of single-particle QSH state at charge neutrality}

\begin{figure*}
\includegraphics[width=7in]{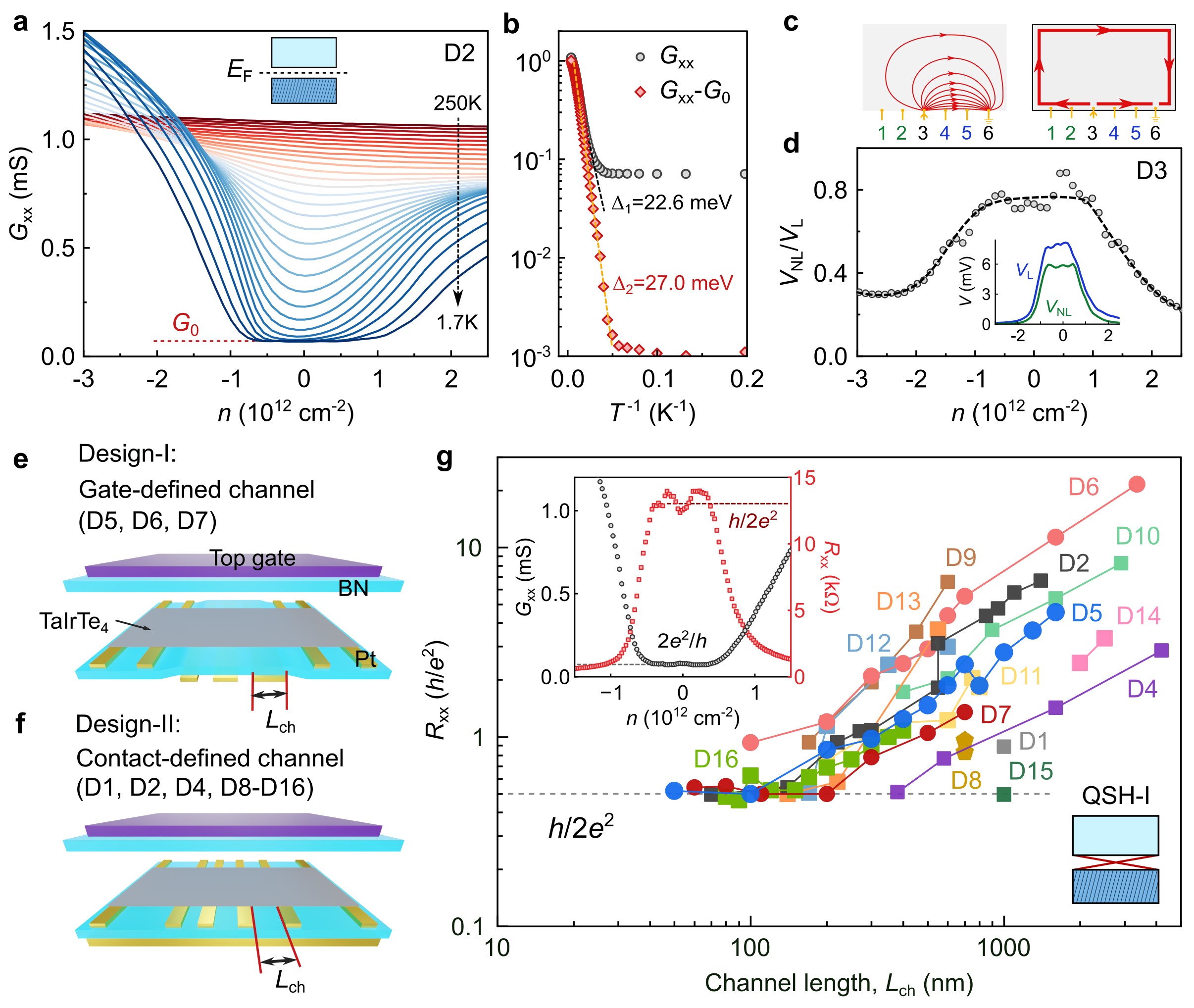}
\caption{{\figtitle{Quantum spin hall (QSH) edge conduction at the charge neutrality in monolayer TaIrTe$_4$.}
\textbf{a,} The four-probe conductance $G_\mathrm{xx}$ of device D2 with a channel length $L_\mathrm{ch}$ = 140 nm at different temperatures. $G_\mathrm{xx}$ plateaued to a constant value $G_0$ at the charge neutrality at low temperatures.  
\textbf{b,} The band gap $\Delta$ is extracted to be $\sim 22.6$ meV (without subtracting $G_0$) and $\sim 27.0$ meV (subtracting $G_0$), determined from the Arrhenius fitting $G_\mathrm{xx}(T) \propto \exp(-\Delta/2k_\mathrm{B}T)$.
\textbf{c,} Schematic of current distribution based on bulk (left) and edge (right) contributions. The bulk (ohmic) contribution results in a weak nonlocal response, while the edge contribution leads to a significant nonlocal voltage drop. 
\textbf{d,} The ratio between nonlocal and local voltages $V_\mathrm{NL}/V_\mathrm{L}$ is large inside the CNP gap measured with device D3 ($L_\mathrm{ch} =$ 1 $\mu$m, also see Extended Data Fig.~\ref{NL_without_edge}).
\textbf{e-f,} Schematic illustrations of two kinds of short-channel devices: gate-defined channel with $L_\mathrm{ch}$ determined by the gate width (\textbf{e}) and contact-defined channel with $L_\mathrm{ch}$ determined by the distance between two adjacent contacts (\textbf{f}). 
\textbf{g,} Channel length dependence of the plateau resistance at CNP. The resistance reaches $h/2e^2$ for $L_\mathrm{ch} \leq$ 100 nm in devices D2 and D5, $L_\mathrm{ch} \leq 200$ nm in device D7, $L_\mathrm{ch}$ = 380 nm (inset) in device D4, and $L_\mathrm{ch} \leq 150$ nm in device D16. For clarity, individual curves for D2, D5, D7, D16 are shown in Extended Data Fig.~\ref{channel_length_seperated}. More data related to panel \textbf{g} can be found in SI 4.1.}}\label{Fig2}
\end{figure*}

We first study the physics associated with the resistance peak at the charge neutrality point (CNP). As shown in Fig.~\ref{Fig2}\textbf{a}, the CNP conductance decreases with decreasing temperature and expands into a plateau below 30 K, suggesting a band gap. Also, the conductance saturates at a nonzero value denoted as $G_0$ in Fig.~\ref{Fig2}\textbf{a}, hinting at an additional in-gap conduction channel. Importantly, the conductance of this plateau is $G_0 \sim 72.8$ $\mu$S, approaching $2e^2/h$ ($\sim 77.5$ $\mu$S). To estimate the magnitude of the band gap, we apply the thermal-activation fit to the temperature-dependent conductance at CNP (with and without subtracting $G_0$) (Fig.~\ref{Fig2}\textbf{b}). The fitting yields an activation gap between 20 to 30 meV, which approximately aligns with the first-principles calculated gap at the CNP (Fig.~\ref{Fig1}\textbf{c}). In SI 1.4, we further extracted the gap size on devices in which the edge channels are physically covered.\\

The finite conductance plateau at the CNP suggests in-gap states, which could arise from edge states or bulk defect states. To gain further insight, we performed nonlocal transport measurements, previously proven effective in studying QSH systems~\cite{roth2009nonlocal, fei2017edge} in which significant nonlocal voltage, $V_\mathrm{NL}$, arises from the edge conduction around the outer boundary of the sample (Fig.~\ref{Fig2}\textbf{c}). In our measurements, $V_\mathrm{NL}$ is prominent within the band gap and suppressed outside the gap (Fig.~\ref{Fig2}\textbf{d}). Moreover, we found negligible $V_\mathrm{NL}$ when the edge contribution was eliminated by physically covering the edge with an insulating boron nitride (BN) flake (Extended Data Fig.~\ref{NL_without_edge}). These observations suggest that the finite conductance plateau in Fig.~\ref{Fig2}\textbf{a} originates from edge state conduction.\\

We now study the nature of the edge states (topological or trivial). A defining characteristic of QSH is the quantized conductance of $2e^2/h$ ($e^2/h$ per edge) in the short-channel (ballistic) regime~\cite{konig2007quantum, du2015robust, wu2018observation, zhao2024realization}. Therefore, we now investigate the edge conduction versus the channel length $L_\mathrm{ch}$ via two device designs. In Design-I, a strategy previously utilized in WTe$_2$~\cite{wu2018observation}, we apply the global top gate to heavily dope the sample and adjust the local bottom gate to neutralize the charge, effectively setting $L_\mathrm{ch}$ as the width of the designated local bottom gate (Fig.~\ref{Fig2}\textbf{e}). In Design-II, the channel is more directly defined between adjacent metal contacts (Fig.~\ref{Fig2}\textbf{f}). Both designs yielded consistent results in our case: the resistance (conductance) decreases (increases) as $L_\mathrm{ch}$ decreases. Upon entering the short-channel regime, the resistance (conductance) approaches and levels off at the quantized value of $h/2e^2$ ($2e^2/h$) regardless of the channel width (Fig.~\ref{Fig2}\textbf{g}). The edge quantization exhibits minimal dependence on current biasing within 100 nA (SI 2.2), suggesting ohmic contacts to the edge channels. The quantization channel length typically ranges from 100 to 200 nm, with our highest-performing sample extending up to 380 nm. Beyond this length, conductance may still saturate to a value on the order of $2e^2/h$ below 30 K (Extended Data Fig.~\ref{extended_gap_fitting}), implying minimal bulk contribution. This suggests that deviations from quantization are likely due to inelastic edge scattering~\cite{xu2006stability,maciejko2009kondo} rather than bulk residual channels.\\ 

We conducted further investigations into the magnetic field $B$ dependence of the edge conduction~\cite{zhang2014robustness, ma2015unexpected}. Our data show that the edge conduction is weakly suppressed by applying magnetic field $B$ (Extended Data Fig.~\ref{g_factor}). We propose an explanation based on our DFT calculation shown in Fig.~\ref{Fig1}\textbf{c}: the edge Dirac point resides within the valence band of the bulk, making it impossible to position the chemical potential solely within the magnetic gap, leading to a weak magnetic field response (See SI 2.5 for more detailed discussions).

\vspace{-4 mm}
\subsection*{Unexpected insulating states at finite dopings} 

\begin{figure*}
\includegraphics[width=7in]{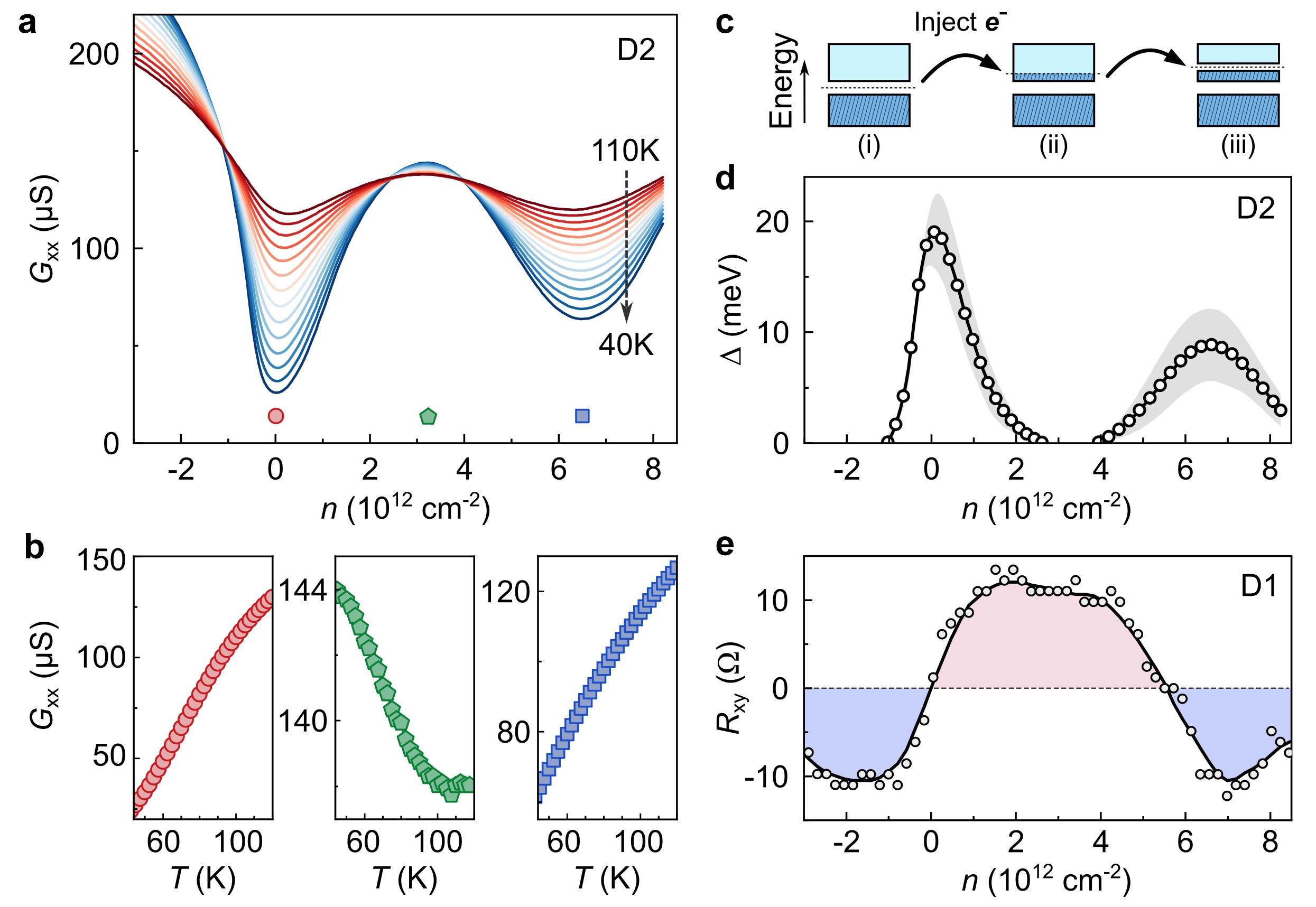}
\caption{{\figtitle{Emergent insulating state at finite doping.}
\textbf{a,} The temperature-dependent four probe conductance $G_\mathrm{xx}$ versus carrier density $n$ across a temperature range from 110 K to 40 K (device D2,  $L_{\mathrm{ch}}= 550$ nm). 
\textbf{b,} Temperature dependence behavior at three representative carrier densities. The colors of the data points in \textbf{(b)} correspond to the carrier densities marked in (\textbf{a}).
\textbf{c,} Sketch of a scenario of correlation-induced gap opening at finite doping. 
\textbf{d,} The size of the fitted activation gap versus $n$. The shaded region in grey delineates the error bounds associated with the energy gap fits.
\textbf{e,} The anti-symmetrized Hall resistance $R_\mathrm{xy}$ versus $n$ at $B=5$ T and $T=60$ K (see SI 3.1 for details). 
}}\label{Fig3}
\end{figure*}

We now turn our attention to investigating the resistance peak in the electron-doped regime, which contradicts the expectations based on the single-particle band structure. Extrinsically, this second resistance peak may be due to disorder-induced localization~\cite{kramer1993localization}, while intrinsically, it may arise from the opening of an energy gap at a specific chemical potential due to electron correlations. To explore this further, we present the temperature-dependent conductance across a wider range of carrier densities in Fig.~\ref{Fig3}\textbf{a} as well as in Extended data Fig.~\ref{Tdep}. At CNP ($n=0$), the temperature dependence displays insulating behavior (Fig.~\ref{Fig3}\textbf{b}, red curve), consistent with the CNP single-particle gap. As we increase the carrier density to $n=3.2\times 10^{12}$ cm$^{-2}$, the temperature dependence becomes metallic (green curve). Further increasing to $n=6.5\times 10^{12}$ cm$^{-2}$ results in insulating behavior once again (blue curve). In a typical scenario, if disorder played a significant role in inducing insulating behavior at $n=6.5\times 10^{12}$ cm$^{-2}$, it should similarly result in insulating behavior at lower carrier densities such as $n=3.2\times 10^{12}$ cm$^{-2}$. Figure~\ref{Fig3}\textbf{d} shows the fitted activation gap $\Delta$ as a function of carrier density $n$.\\

We also performed Hall measurements at an elevated temperature of 60 K. We observed not only the expected sign reversal of the Hall signal at the CNP but also a second sign reversal near the resistive feature at $6.5\times 10^{12}$ cm$^{-2}$ (Fig.~\ref{Fig3}\textbf{e}). This additional sign reversal could arise from the emergence of hole-like pockets, consistent with the existence of VHS (Fig.~\ref{Fig1}\textbf{c} and additional data and analysis in SI 3.1-3.3). Collectively, we propose the following explanation for the double resistance peaks observed in Fig.~\ref{Fig1}\textbf{f}: The first resistance peak at CNP represents a single-particle gap. Upon electron doping to a certain level, a correlated gap appears from enhanced correlations near the VHS, as illustrated in Fig.~\ref{Fig3}\textbf{c}.

\vspace{-4 mm}
\subsection*{Observation of a second QSH state - TaIrTe$_4$ is a dual QSH insulator}

\begin{figure*}
\includegraphics[width=3.5in]{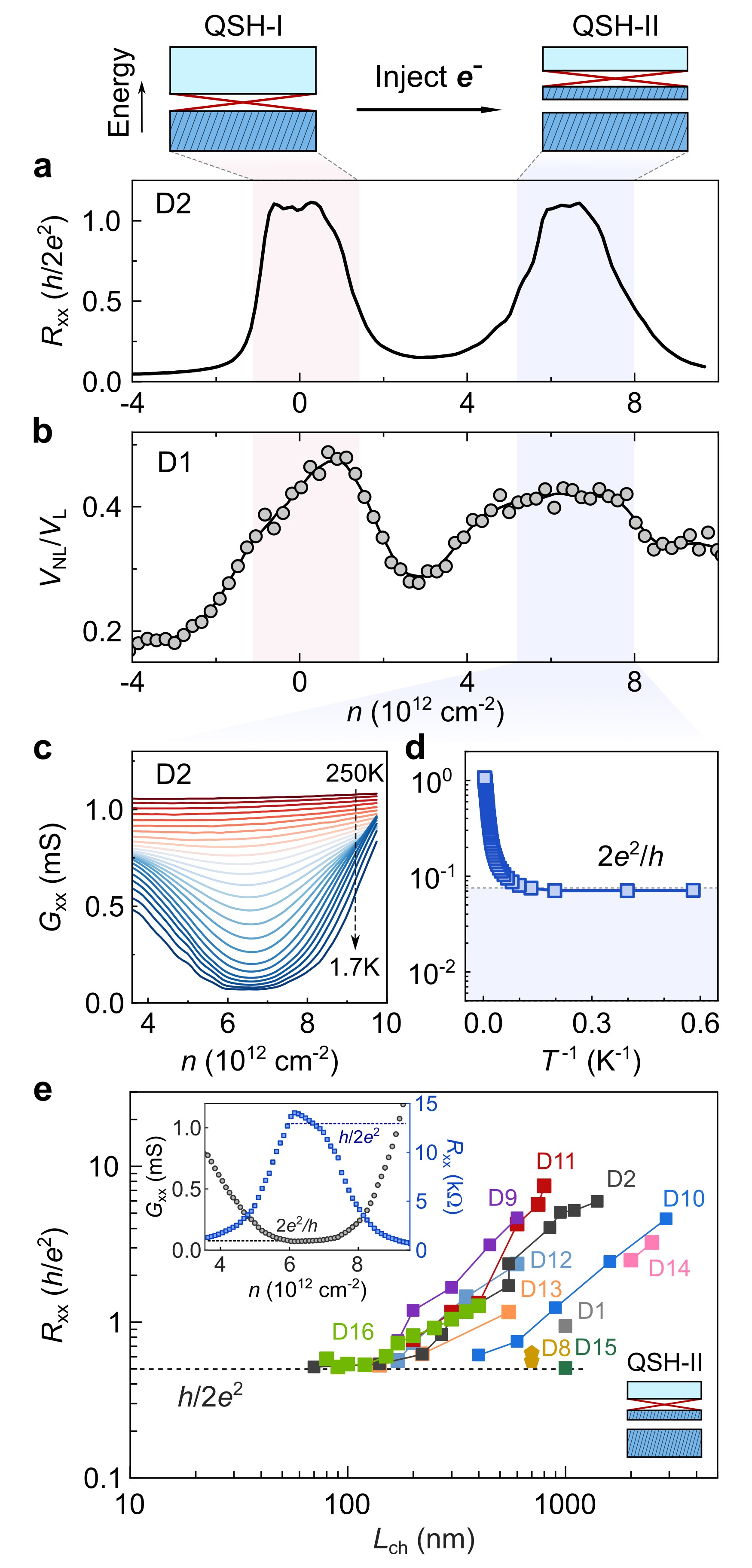}
\caption{{\figtitle{The dual quantum spin hall (dual-QSH) effect in monolayer TaIrTe$_4$.}
\textbf{a,} Resistance versus carrier density (D2, $L_{\mathrm{ch}} = 140$ nm) at $T$ = $1.7$ K and the schematic of the dual-QSH. Two resistance plateaus are developed at CNP and $n \sim 6.5 \times 10^{12}$ cm$^{-2}$, corresponding to the QSH-I and QSH-II states, respectively. 
\textbf{b,} The ratio between nonlocal and local voltages ($V_\mathrm{NL}/V_\mathrm{L}$) has two maxima around the CNP and $n \sim 6.5 \times 10^{12}$ cm$^{-2}$ (measured in device D1, $L_\mathrm{ch} = 1$ $\mu$m, also see Extended Data Fig.~\ref{NL_eside}).  
\textbf{c,} Conductance versus $n$ around the emergent insulating gap (D2, $L_{\mathrm{ch}}= 140$ nm).  
\textbf{d,} The minimum conductance in (\textbf{c}) versus temperature ($T^{-1}$), which plateaus at $\sim2e^2/h$ at low temperatures. 
\textbf{e,} Channel length dependence of the resistance at the emergent gap. The resistance approaches $\sim h/2e^2$ for $L_\mathrm{ch} \leq 140$ nm in device D2 (data for $L_{\mathrm{ch}}= 70$ nm is shown in the inset), and $L_\mathrm{ch} \leq 120$ nm in device D16. For clarity, individual curves for D2, D16 are shown in Extended Data Fig.~\ref{channel_length_seperated} with more data in SI 4.1.}}\label{Fig4}
\end{figure*}

Before moving to more in-depth theoretical considerations, we now explore the topological nature of the emergent insulating state. This is particularly interesting since the new insulating gap occurs at a low filling ($\sim6.5\times10^{12}$ cm$^{-2}$) of the QSH conduction band. Our initial findings revealed that the ratio between $V_\mathrm{NL}$ and $V_\mathrm{L}$ is significant within both the CNP single-particle gap and the second insulating gap while suppressed elsewhere (Fig.~\ref{Fig4}\textbf{a-b}). This indicates that edge conduction is present within the second gap as well.\\

To definitively discern if the second insulating state is topological or trivial, we proceeded to measure the conductance of short-channel devices. In Fig.~\ref{Fig4}\textbf{c-d}, we investigate the temperature-dependent conductance of a fixed short channel length ($140$ nm) in the vicinity of the second insulating gap. As the temperature decreases, we observe that the conductance stabilizes at a plateau of $2e^2/h$. We further investigate the channel length dependence of the conductance at the second gap. As shown in Fig.~\ref{Fig4}\textbf{e}, the resistance (conductance) decreases (increases) with decreasing the channel length $L_\mathrm{ch}$. Upon entering the short-channel regime, the resistance (conductance) levels off at the quantized value of $h/2e^2$ ($2e^2/h$). We also examined the dependence on the magnetic field, which shows a conductance suppression by magnetic field similar to that for the CNP state (SI 4.3). In conclusion, we establish that the second insulating gap also exhibits the QSH state with quantized helical edge conduction.\\ 

Beyond the quantized conduction, we have further conducted spin valve experiments~\cite{li2014electrical}. Our results, provided in SI 4.5, indicate that the edge currents within both insulating gaps carry spin. Collectively, we provide strong evidence that monolayer TaIrTe$_4$ is a novel \textit{dual} QSH insulator, which consists of two QSH states, both featuring the hallmark helical quantized edge conduction (Fig.~\ref{Fig4}\textbf{a}): one is the consequence of the single-particle topology at CNP and the other may arise from enhanced correlations at finite dopings. The two QSH states can be switched back and forth by tuning the carrier density via gating.

\vspace{-4 mm}
\subsection*{Theoretical understanding of the dual QSH state} 

We now provide theoretical analyses and present a tentative framework to understand the observed dual QSH insulator state. We outline our theoretical understanding in a step-by-step manner as follows.\\
 
First, we study the single-particle band structure to understand the QSH at charge neutrality. Our first-principles calculations have revealed that the low-energy electronic states primarily originate from the Ta 5$d$ orbitals (see Extended Data Fig.~\ref{orbital} and SI 5.1). The presence of two inequivalent Ta atoms results in two distinct bands, and the fact that they are related by the $C_{2a}$ symmetry implies that these two bands will exhibit Dirac crossings (see Extended Data Fig.~\ref{band_inversion}). Indeed, our calculations on monolayer TaIrTe$_4$ show two Dirac cones at the $\Lambda$ points without spin-orbit coupling (SOC). The inclusion of SOC gaps the Dirac point, giving rise to the QSH insulating state (see SI 5.1).\\

Second, our calculations revealed saddle-point VHS inside the conduction (valence) band. Monolayer TaIrTe$_4$ has three low-energy Fermi pockets with direct gaps (Fig.~\ref{Fig5}\textbf{a}) - Two arise from the gapped Dirac cones at $\Lambda$, while the third comes from an additional pocket at $\Gamma$. As one increases energy, these pockets are found to gradually expand and merge into saddle points in the conduction and valence bands, giving rise to VHS (Figs.~\ref{Fig5}\textbf{a-c}). Interestingly, our calculations indicate that achieving the VHS on the hole side requires a higher carrier density. Consequently, we might also expect to observe a VHS-induced correlated insulating peak on the hole side, but this would necessitate reaching higher gate voltages on the hole side. Currently, our data only shows a slight increase in resistance on the hole side, as shown in Fig.~\ref{Fig1}\textbf{f}.\\

Third, we attempt to understand the emergent insulating state. Based on the carrier density, the absence of evidence for magnetism, and the value of quantized edge conductance associated with the correlated insulating gap, we identify charge density wave (CDW) with translational symmetry breaking as a reasonable state (see detailed discussion in SI 6.1). We computed the electronic charge susceptibility at various wavevectors (Q) via the Lindhard function (Methods). The calculated charge susceptibility shows multiple distinct local maxima at different $Q$ vectors (Fig.~\ref{Fig5}\textbf{d}). Among these, we identify a specific local maximum at $Q^*=(0.068\frac{2\pi}{a},0\frac{2\pi}{b})$ for the following reasons: (1) It aligns with the 1D Ta chain direction (Fig.~\ref{Fig5}\textbf{d} left inset). (2) It corresponds to a relatively large supercell spanning 15 atomic unit cells ($\sim$ 5.6 nm), consistent with the low density of the second gap. (3) It connects two adjacent VHS (Fig.~\ref{Fig5}\textbf{a}) at the energy where the Fermi surface exhibits two parallel contours (\textcircled{3} in Fig.~\ref{Fig5}\textbf{c}). Subsequently, we computed the folded band structure in the $Q^*$ defined superlattice (Fig.~\ref{Fig5}\textbf{e-f} and Methods). Indeed, the introduction of the $Q^*$ superlattice potential results in the opening of an insulating gap (Fig.~\ref{Fig5}\textbf{f}), providing a plausible explanation for our experimental data.\\

Fourth, we continue with the CDW scenario and theoretically evaluate the topology of the CDW gap. As shown in Fig.~\ref{Fig5}\textbf{f}, we found that the CDW gap indeed hosts nontrivial QSH states by calculating the $\mathcal{Z}_2$ invariant via the Wilson loop (Methods). Our calculations interestingly show that the nature of the CDW - whether it is topological or trivial - is closely tied to the periodicity of the CDW, as detailed in SI 6.2. The determined $Q^*$ above leads to a topological CDW whereas larger $Q$ could yield trivial CDW. In contrast, when the CDW periodicity is fixed by $Q^*$, the CDW consistently exhibits a topological nature, regardless of the strength of CDW superlattice potential (SI 6.3).

\vspace{-4 mm}
\subsection*{Discussions}

\begin{figure*}
\includegraphics[width=6.5in]{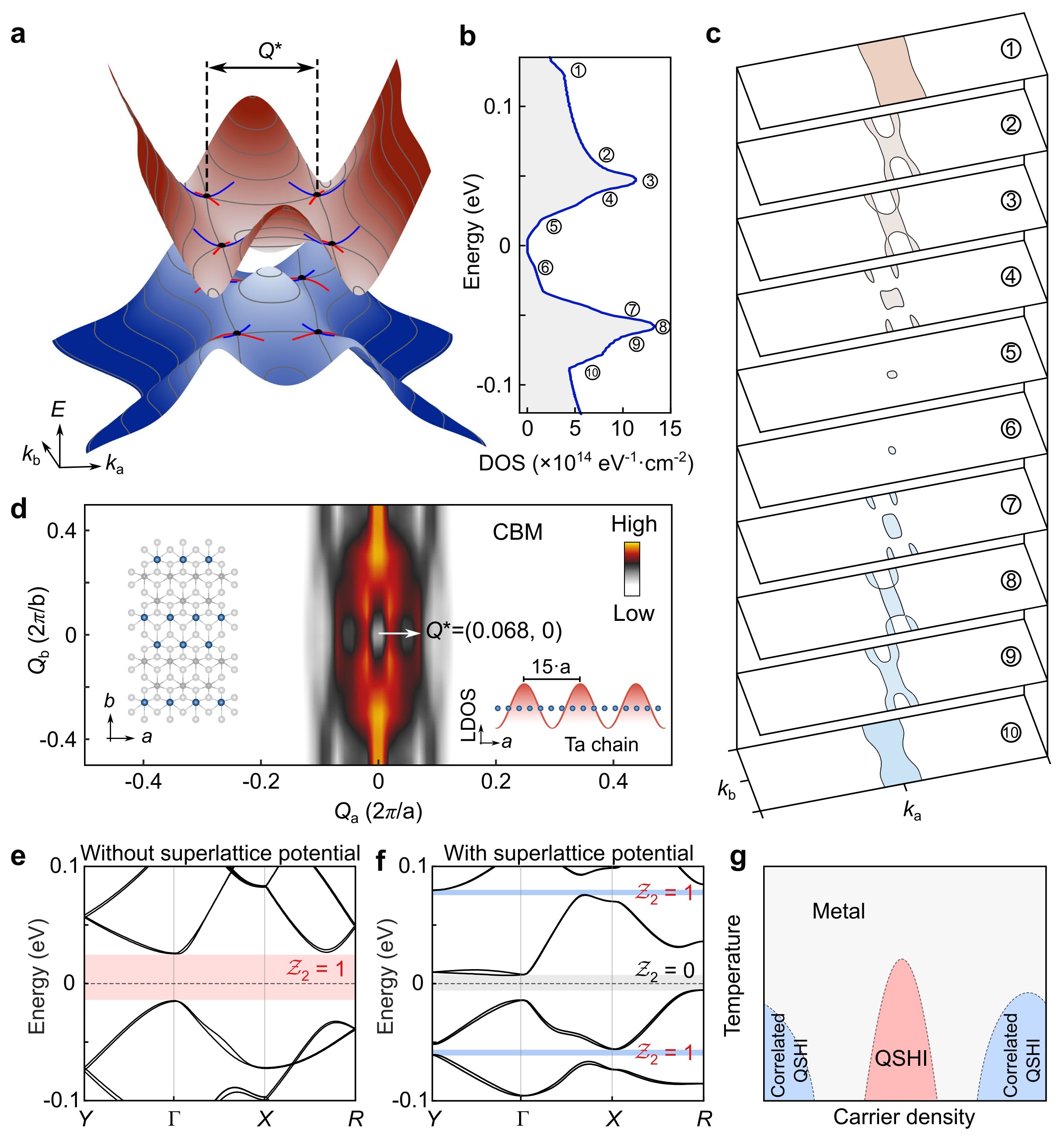}
\caption{{\figtitle{Theoretical investigation of the dual QSH state in monolayer TaIrTe$_4$.}
\textbf{a,} The band structure of monolayer TaIrTe$_4$ with VHS denoted by black dots. 
\textbf{b,} The calculated DOS as a function of energy. 
\textbf{c,} The constant energy contours at different energies, numbered from \textcircled{1} to \textcircled{10}. Their energy values can be found in panel \textbf{(b)}. 
\textbf{d,} Calculated electronic susceptibility of the conduction band based on the first-principles band structure. The white arrow highlights a  local maximum of the electronic susceptibility at a wavevector $Q^*$. $Q^*$ connects two neighboring VHS along the $k_\mathrm{a}$ direction as marked in \textbf{(a)}, which corresponds a superlattice of $\sim15$ unit cells alone the 1D chain direction (see right inset). The left inset shows the atomic structure of TaIrTe$_4$ with Ta chains highlighted. The same calculation for the valence band can be found in SI 5.5. 
\textbf{e-f,} The folded band structure with a superlattice periodicity of 15 unit cells,  \textbf{(e)} without superlattice potential, and \textbf{(f)} with superlattice potential. 
\textbf{g,} Conjectured phase diagram of monolayer TaIrTe$_4$.
}}\label{Fig5}
\end{figure*}

Our research presents experimental confirmation of a unique dual QSH state in an intrinsic monolayer TaIrTe$_4$. This dual QSH state combines a high-quality single-particle QSH with a novel correlated QSH. These results position TaIrTe$_4$ as an exemplary system for investigating time-reversal-symmetric (fractionalized) topological states and the interplay between topology and the potential charge order (Fig.~\ref{Fig5}\textbf{g}). Employing scanning tunneling microscopy (STM) measurements could shed more light on the nature of the correlated state, as STM is an ideal tool to detect potential density wave orders~\cite{teng2022discovery}. Furthermore, the emergent insulating state might give rise to the spontaneous breaking of various symmetries (e.g., space-inversion, rotation), which, coupled with the significant Berry curvature of the topological bands, could yield strong second-order nonlinear responses at low frequencies~\cite{xu2018electrically,ma2019observation}. Additionally, it's worthwhile to explore the impact of strain, which could influence the VHS and, consequently, the correlated state. Moreover, it is interesting to study bilayer TaIrTe$_4$ which is also predicted to be a QSH insulator~\cite{guo2020quantum}, in sharp contrast to WTe$_2$ where the bilayer becomes a trivial insulator.\\

Our data implies the existence of emergent flat QSH bands: The ``QSH'' aspect is supported by our observation of quantized edge conduction within the emergent insulating gap; the ``flatness'' can be inferred from the low carrier density ($\sim6.5\times 10^{12}$ cm$^{-2}$) associated with the emergent gap. The creation of these flat bands, likely originating from the CDW superlattice with a periodicity of a few nanometers, parallels the extensively studied moiré flat bands. In the future, micro-ARPES measurements with gating capabilities would be valuable for directly resolving the emergent band structures. Additionally, exploring the fractional filling of these bands at ultralow temperatures in high-quality samples holds promise for realizing time-reversal-symmetric fractionalization and topological order~\cite{levin2009fractional, park2010dirac, maciejko2010fractional, santos2011time, goerbig2012fractional,li2014fractional, wang2016time, barkeshli2019symmetry,  wang2022fractional, wu2023time}.

\newpage

\clearpage
\newpage

%\def\refnamex{References}
%\renewcommand\bibfont{\normalsize}
%\putbib
%\end{bibunit}

\section*{Methods}
%\begin{bibunit}
%\newbibstartnumber{60}
\noindent\textbf{TaIrTe$_4$ crystal growth:} TaIrTe$_4$ crystals were synthesized by a solution-growth method with Te flux. High-purity ($> 99.99 \%$) Ta, Ir, and Te elements were mixed at a molar ratio of 3:3:94 in an alumina crucible and sealed in a vacuum quartz tube. The quartz tube was slowly heated up to 1373 K, maintained for 10 hours and then gradually cooled down to 873 K at a rate of 2 K/h. The shiny, needle-like single crystals were separated from the Te flux by centrifuging. The TaIrTe$_4$ single crystals were characterized by atomic-resolution cross-sectional scanning transmission electron microscopy (STEM), selected area electron diffraction (SAED), energy-dispersive X-ray spectroscopy (EDS) and residual resistivity ratio (RRR) for crystalline properties, atomic structure, elemental ratio, and sample electrical quality (see detailed characterizations in SI 1.1-1.2). \\

\noindent\textbf{Mechanical exfoliation and layer number identification of TaIrTe$_4$:} The TaIrTe$_4$ were exfoliated on Si/SiO$_2$ substrates (SiO$_2$ is 285 nm thickness) by the Scotch tape method. To address the air-sensitive nature of thin TaIrTe$_4$ flakes, the exfoliation and the following device fabrication processes were performed in an argon-glovebox. We introduced gentle oxygen plasma treatment and heating process to increase the exfoliation yield. Prior to the exfoliation process, we employed a mild oxygen plasma treatment using the Anatech Barrel plasma cleaning system. The treatment involved a flow rate of 40 sccm at a power setting of 100 W for 10 minutes to clean the Si/SiO$_2$ substrate. Subsequently, we placed the Si/SiO$_2$ substrates onto the TaIrTe$_4$ tape, carefully pressing the stack to eliminate any trapped air bubbles. The tape (with Si/SiO$_2$ substrates) was then heated on a hot plate at a temperature of $80\degree$C for 10 minutes. Finally, we carefully removed the substrates from the TaIrTe$_4$ tape and inspected them for the presence of thin flakes using an optical microscope.\\

We used multiple characterizations to determine the layer number of TaIrTe$_4$ flakes as shown in Extended Data Fig.~\ref{layernumber}, including optical contrast, atomic force microscopy (AFM), Raman spectroscopy, and cross-sectional STEM. Extended Data Fig.~\ref{layernumber}\textbf{a} displays optical images of representative TaIrTe$_4$ flakes situated on a SiO$_2$ substrate, revealing distinct optical contrast variations corresponding to different layer numbers. The quantitative relationship between optical contrast ($C=\frac{I_{\textrm{flake}}-I_{\textrm{substrate}}}{I_{\textrm{flake}}+I_{\textrm{substrate}}}$) and layer numbers can be found in SI 1.2. Extended Data Fig.~\ref{layernumber}\textbf{b} displays the surface morphology of exfoliated TaIrTe$_4$ flakes, examined using a Park NX10 atomic force microscopy (AFM) system in tapping mode with a NCHR tip. The height linecut (inset) reveals a step of approximately 0.8 nm, indicative of a monolayer thickness. Raman spectroscopy also exhibits layer-dependent behavior (Extended Data Fig.~\ref{layernumber}\textbf{c}): with increasing layer numbers, the intensity ratio of A$_2$ (@125.7 cm$^{-1}$ for 1L)/A$_2$ (@137.3 cm$^{-1}$ for 1L) rises, and the A$_1$ mode (@231.8 cm$^{-1}$ for 1L) shifts toward lower wavenumbers. Additionally, cross-sectional STEM results in Extended Data Fig.~\ref{layernumber}\textbf{d} directly confirm the layer numbers (layer thickness approximately 0.7 nm, consistent with AFM results). These diverse characterization methods, collectively, and unambiguously determine layer numbers.\\

\noindent\textbf{Device fabrication and measurements:} Because of the air-sensitivity of thin TaIrTe$_4$ flakes, we adopted the bottom contact structure with BN encapsulation. The metal electrodes (both gates and contacts) were patterned by standard UV or e-beam lithography via the Heidelberg $\mu$PG101 laser writing and the Elionix HS50 e-beam writing systems. Metal electrodes, consisting of 2 nm Ti and 10 nm Pt layers, were deposited using an e-beam evaporator (Angstrom Engineering) and subsequently annealed in a tube furnace at $350 \degree $C for 3 hours in an Ar/H$_2$ (1:1) forming gas atmosphere. The metal electrodes were fabricated under ambient conditions and subsequently transferred into a glovebox for further processing. Within the glovebox, we conducted the assembly of the BN/few-layer graphene stack (for the top gate structure), as well as the exfoliation, identification, stacking, and encapsulation of TaIrTe$_4$ flakes. This glovebox is equipped with a dry transfer stage and an optical microscope to facilitate these operations. For a detailed fabrication process, please refer to Extended Data Fig.~\ref{TIT_fab} and SI 1.3. To safeguard the final device, we applied a PMMA capping layer via spin coating while still within the glovebox.\\

Electrical transport measurements were conducted using a variety of cryogenic systems, including the Montana cryostat, Quantum Design Opticool, PPMS, and MPMS3 systems. Standard lock-in techniques were predominantly employed to measure most of the electrical signals. Gate voltages were applied using Keithley source meters. In dual-gated devices, the charge carrier densities ($n$) and the external electric field ($D$) can be independently adjusted through the combination of the top gate voltage (V$_\mathrm{tg}$) and the bottom gate voltage ($V_\mathrm{bg}$).
\begin{equation}
n=\frac{\epsilon_{\textrm{0}}\epsilon_{\textrm{BN}}V_{\textrm{bg}}}{ed_{\textrm{b}}}+\frac{\epsilon_{\textrm{0}}\epsilon_{\textrm{BN}}V_{\textrm{tg}}}{ed_{\textrm{t}}},
\label{n_equation}
\end{equation}
\begin{equation}
D=\frac{1}{2}(\frac{\epsilon_{\textrm{0}}\epsilon_{\textrm{BN}}V_{\textrm{bg}}}{d_{\textrm{b}}}-\frac{\epsilon_{\textrm{0}}\epsilon_{\textrm{BN}}V_{\textrm{tg}}}{d_{\textrm{t}}}),
\label{E_equation}
\end{equation}
Here, $\epsilon_{\textrm{0}}$ is the vacuum permittivity, $\epsilon_{\textrm{BN}}$ is the BN dielectric constant ($\epsilon_{\textrm{BN}}\approx 3$) and $d_\textrm{b}(d_\textrm{t})$ is the thickness of the bottom (top) BN dielectric.\\

\noindent\textbf{First-principles calculations:} 
The electronic structures of TaIrTe$_4$ were calculated from density functional theory via the Vienna ab-initio simulation package (VASP) \cite{kresse1996efficient}. The exchange correlation functional was considered by the generalized gradient approximation (GGA) \cite{perdew1996generalized}. The energy cutoff of plane waves was set to be $500$ eV and the smearing width of the Gaussian smearing method was chosen to be $0.05$ eV. A Gamma-centered $k$ mesh ($16\times8\times 1$) was used for the BZ integrations.  To get the ground state, the convergence criteria of lattice optimizations were set to 10$^{-6}$ eV and 0.1 meV/$\AA$ for the total energy and ionic forces, respectively. The optB88-vdW correlation functional~\cite{klimevs2011van} was used to describe the vdW interactions for 2D materials. The obtained Bloch wavefunctions were projected onto Wannier functions via using the WANNIER90 package~\cite{mostofi2008wannier90}.  The density functional perturbation theory implemented in VASP was performed to obtain the force constants with a $3\times2\times1$ supercell for phonon dispersion by Phonopy code~\cite{togo2015first}. The real space Hamiltonian $H_\mathbf{r}$ was constructed by $s$ and $d$ orbitals of Ta, $p$ orbital of Te, $s$ and $d$ orbitals of Ir. The edge states were calculated by iterating the surface Green function \cite{bryant1985surface}. The electronic susceptibility $\chi(q)$ was calculated by the Lindhard function:
\begin{equation}
\chi(\mathbf{q})=\sum_{\mathbf{k}}\frac{f_{\mathbf{k}}(1-f_{\mathbf{k}+\mathbf{q}})}{\varepsilon_{\mathbf{k}+\mathbf{q}}-\varepsilon_{\mathbf{k}}+i\Gamma}
\end{equation}
where $f$ is the Fermi Dirac distribution, $\varepsilon$ is the momentum-dependence energy and $\Gamma$ is the broadening. 
The CDW modulation of electronic structure was performed by the Fr\"{o}hlich-Peierls model \cite{frohlich1954theory}:
\begin{equation}
H=H_{\mathbf{r}}+V\cos(Q\cdot x)\psi_{\mathbf{r}}^{\dagger}\psi_{\mathbf{r}}
\end{equation}
where $V$ is the modulation strength and only the intra-orbital interaction is considered. The periodicity of potential was determined by the $\mathbf{Q}^*$ vector denoted in Fig.~\ref{Fig5}\textbf{d}.\\

To figure out the origin of the topology of the band with superlattice potential, we perform calculations of bands and Wilson loop with continuously changed periodicity based on an eight-band tight-binding model. The eight-band tight-binding model is fitted by WANNIER90 package~\cite{mostofi2008wannier90} and constructed by two Ta's $d_{xz}$-orbital and two Te's $p_z$-orbital with spin-up, down components and spin-orbit coupling. We confirmed that our eight-band tight-binding model captures the same saddle points and helical edge as the tight-binding model with full orbitals.\\

\textbf{Acknowledgements:} 
We thank Jennifer Cano, Valla Fatemi, Rafael Fernando, Luqiao Liu, Haizhou Lu, Xiao-Bin Qiang, Ying Ran, Brian Skinner, Aviram Uri, Ilija Zeljkovic, Fan Zhang, and Yi Zhang for fruitful discussions. Q.M. acknowledges the support by the AFOSR grant FA9550-22-1-0270 (transport measurements and data analysis). Q.M. and S.-Y.X. acknowledge the support from the Center for the Advancement of Topological Semimetals, an Energy Frontier Research Center funded by the US Department of Energy Office of Science, through the Ames Laboratory under contract DE-AC02-07CH11358 (device fabrication). Q.M. also acknowledges support from the NSF CAREER award DMR-2143426 (manuscript writing), the CIFAR global scholar program, and the Alfred P. Sloan Foundation. G.C. acknowledges the support from the National Research Foundation, Singapore under its Fellowship Award (NRF-NRFF13-2021-0010) and the Nanyang Assistant Professorship grant. N.N. acknowledges the supports from the US DOE, Office of Science, under Award Number DE-SC0021117 (single crystal growth and characterization of TaIrTe$_4$). Y.S. acknowledges the support from the Strategic Priority Research Program of the Chinese Academy of Sciences (grants number XDB33030000) and the Informatization Plan of Chinese Academy of Sciences(CAS-WX2021SF-0102). K.S.B. and Y.W. acknowledges the support by the Air Force Office of Scientific Research under award number FA9550-20-1-0246. X.Q. acknowledges the support by the National Science Foundation (NSF) under Award Number DMR-1753054 and by the donors of ACS Petroleum Research Fund under grant number 65502-ND10. D.C.B. acknowledges the support from Harvard University Center for Nanoscale Systems, a member of the National Nanotechnology Coordinated Infrastructure Network, under NSF award number ECCS-2025158, and the STC Center for Integrated Quantum Materials, NSF grant number DMR-1231319. Z.S. acknowledges the support from Swiss National Science Foundation under grant number P500PT-206914. M.G. acknowledges the support of the National Science Foundation (NSF) EPMD program via grant 2211334. A.S. acknowledges the support by DMR-2103842. Portions of this research were conducted with the advanced computing resources provided by Texas A\&M High Performance Research Computing. J. L. and Y. Z. are partly supported by the National Science Foundation Materials Research Science and Engineering Center program through the UT Knoxville Center for Advanced Materials and Manufacturing (Grant number DMR-2309083). L.F. and Q.M. acknowledge support from the National Science Foundation Convergence Program under grant number ITE-2235945. Ziqiang Wang is supported by the U.S. Department of Energy, Basic Energy Sciences grant number DE-FG02-99ER45747.  K.W. and T.T. acknowledge support from the JSPS KAKENHI (Grant Numbers 21H05233 and 23H02052) and World Premier International Research Center Initiative (WPI), MEXT, Japan. We also acknowledge that some of the work was done in the Boston College cleanroom and nanotechnology facilities.\\

\textbf{Author contributions:} Q.M. conceived the experiments and supervised the project. J.T. fabricated the devices with the help of T.S.D., A.G., M.G., S.-Y.X., and K.S.B. J.T. performed the electrical measurements, and analyzed data with the assistance from T.S.D, A.G., Z.H., Z.S. M.S., V.B. and Z.H.W. A.S. and X.Q. performed the first-principles calculations of the single-particle band; H.C. and G.C. performed the calculations for different edge terminations, electronic susceptibility, CDW band structures, and topology; J.L. and Y.Z. performed the phonon dispersion calculation. L.F. and Z.Q.W. supplied the theory inputs. Y.W. and K.S.B. performed the Raman measurements. D.C.B. carried out the TEM characterization. T.Q., X.H., Y.S., and N.N. grew the TaIrTe$_4$ bulk crystals. K.W. and T.T. grew the BN bulk crystals. Q.M., S.-Y.X., and J.T. wrote the manuscript with input from all authors.\\

\textbf{Competing interests:} The authors declare that they have no competing interests.\\

\textbf{Data availability:} Source data are available at https://doi.org/10.7910/DVN/JJAI19. All other data that support the plots within this paper and other findings of this study are available from the corresponding authors upon reasonable request.

\clearpage

\setcounter{figure}{0}
\renewcommand{\figurename}{Extended Data Figure}

\begin{figure*}
\includegraphics[width=6.8 in]{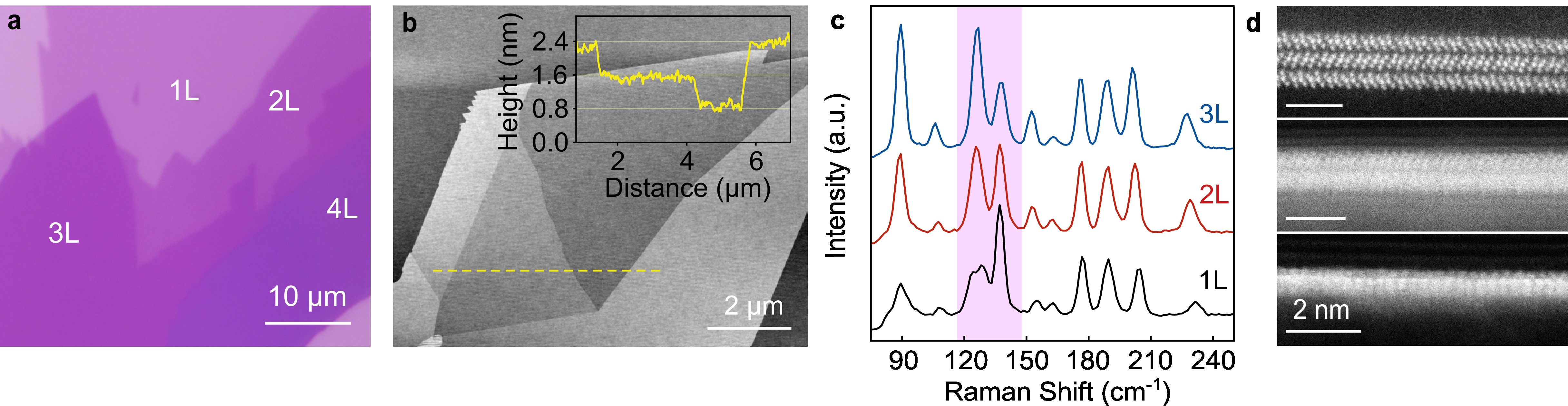}
\caption{{\figtitle{Basic characterizations of TaIrTe$_4$ flakes.}
\textbf{a,} Optical image displaying exfoliated TaIrTe$_4$ flakes on a Si/SiO$_2$ substrate.
\textbf{b,} AFM image of a TaIrTe$_4$ flake. Inset exhibits the linecut profile, indicating a step thickness of approximately 0.8 nm.
\textbf{c,} Raman spectrum of the TaIrTe$_4$ flakes ranging from 1L to 3L. With increasing layer numbers, the intensity ratio of A$_2$ (@125.7 cm$^{-1}$ for 1L)/A$_2$ (@137.3 cm$^{-1}$ for 1L) rises, and the A$_1$ mode (@231.8 cm$^{-1}$ for 1L) shifts toward lower wavenumbers.
\textbf{d,} Cross-sectional STEM images of monolayer, bilayer, and trilayer flakes.
}}\label{layernumber}
\end{figure*}

\begin{figure*}
\includegraphics[width=6.8 in]{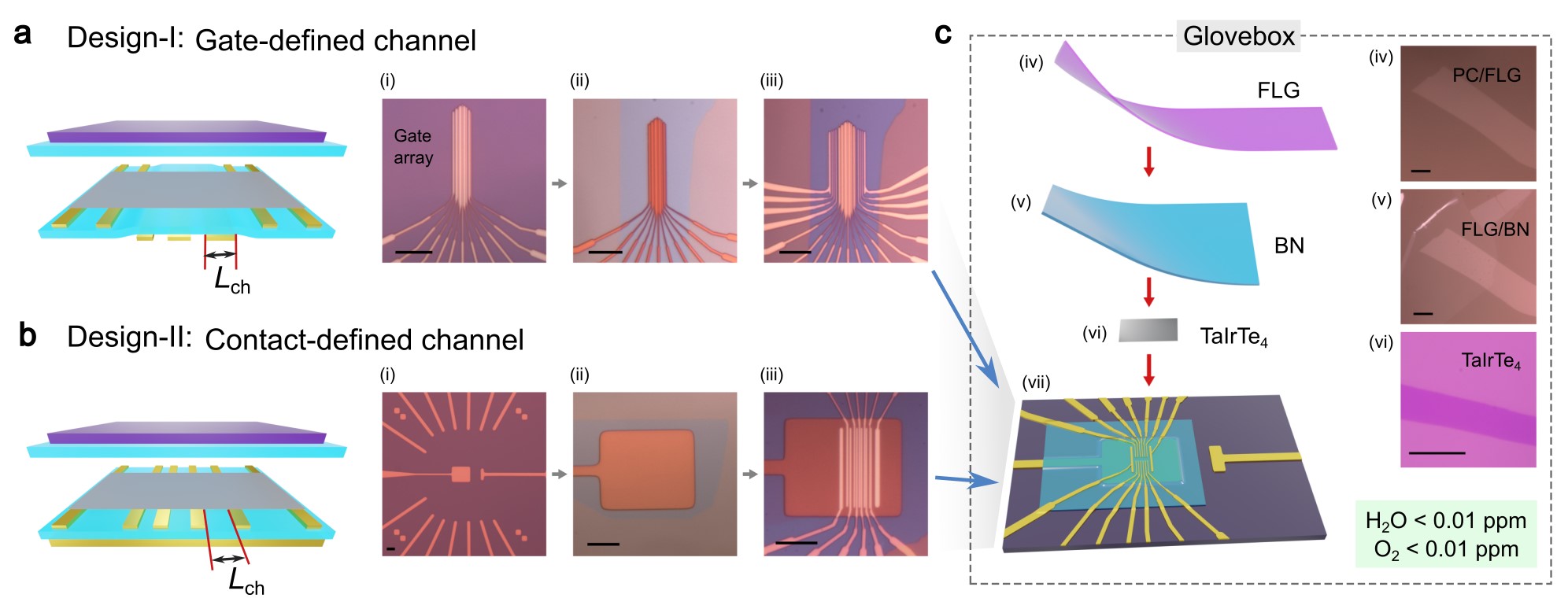}
\caption{{\figtitle{Fabrication processes of TaIrTe$_4$ devices.}
\textbf{a-b,} Schematic and corresponding fabrication processes for Design-I (\textbf{a}) and Design-II (\textbf{b}) bottom structures. \textbf{c,} Fabrication processes of the top structure (involving TaIrTe$_4$ and top gate) conducted inside an argon-glovebox. Scale bars: 10 $\mu$m.}}\label{TIT_fab}
\end{figure*}

\begin{figure*}
\includegraphics[width=6.8 in]{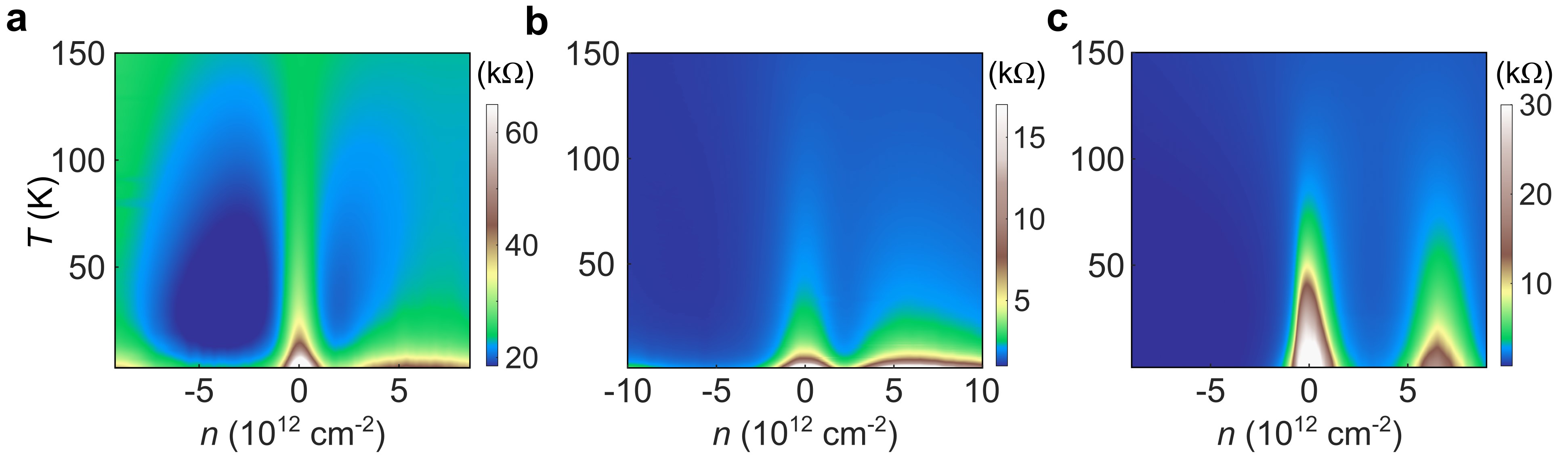}
\caption{{\figtitle{Four-probe resistance versus carrier density and temperature.} \textbf{a,} Device D17; \textbf{b,} Device D1; \textbf{c,} Device D2.}}
\label{Tdep}
\end{figure*}

\begin{figure*}
\includegraphics[width=6.5in]{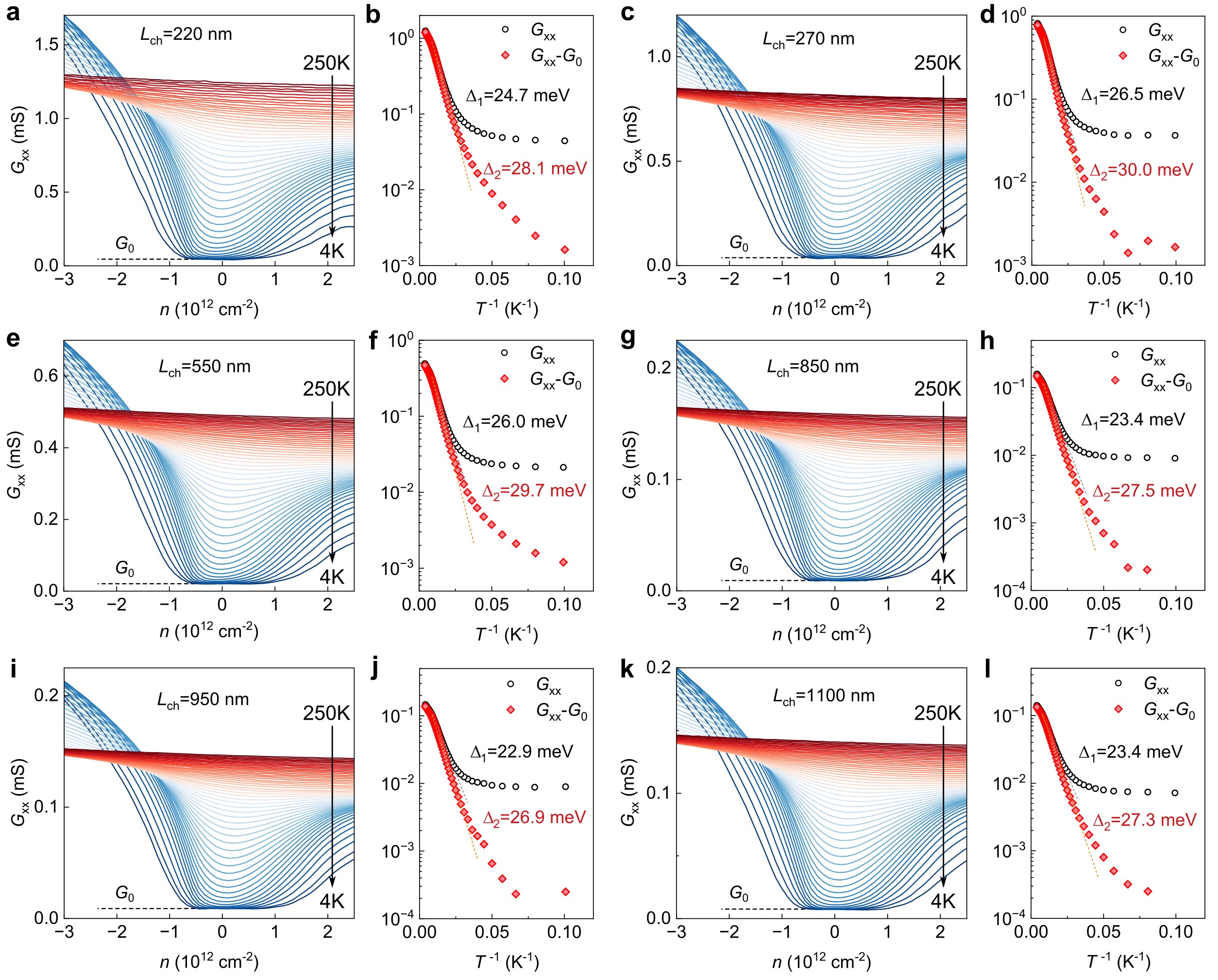}
\caption{{\figtitle{Temperature dependence of CNP conductance and thermal activation fitting (device D2).} The four-probe conductance ($G_\mathrm{xx}$) versus carrier density ($n$) is shown with temperatures ranging from 250 K to 4 K, along with corresponding thermal activation gap fits. Channel lengths: \textbf{a-b,} $L_\mathrm{ch}=220$ nm, \textbf{c-d,} $L_\mathrm{ch}=270$ nm, \textbf{e-f,} $L_\mathrm{ch}=550$ nm, \textbf{g-h,} $L_\mathrm{ch}=850$ nm, \textbf{i-j,} $L_\mathrm{ch}=950$ nm, \textbf{k-l,} $L_\mathrm{ch}=1100$ nm.}}
\label{extended_gap_fitting}
\end{figure*}

\begin{figure*}
\includegraphics[width=6in]{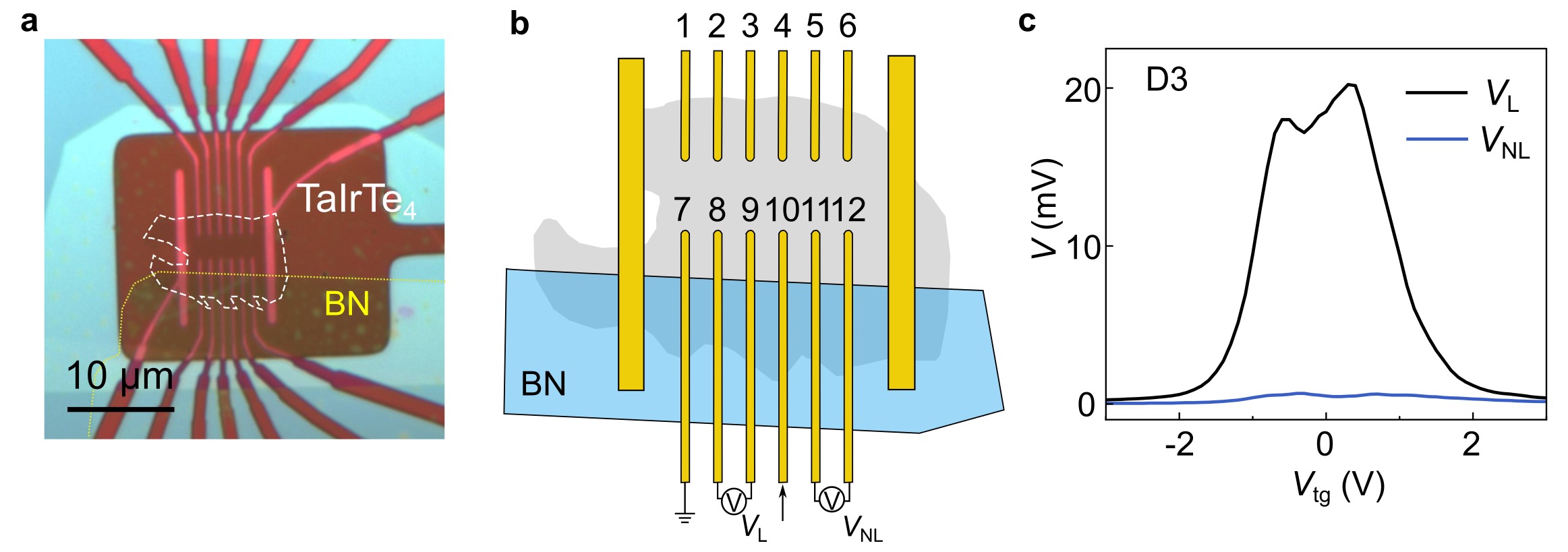}
\caption{{\figtitle{Nonlocal measurements without edge contribution.}
\textbf{a,} An optical image of device D3 with half of its boundaries covered by BN. Scale bar: 10 $\mu$m.
\textbf{b,} Device schematic and contact labeling.
\textbf{c,} Plots of local and nonlocal voltages versus carrier density $n$. The current was injected from Contact 10 to 7 ($I_\mathrm{xx}$ = 100 nA), and the voltages were measured between Contact 9 and 8 ($V_\mathrm{L}$) and between Contact 11 and 12 ($V_\mathrm{NL}$).}}
\label{NL_without_edge}
\end{figure*}

\begin{figure*}
\includegraphics[width=7in]{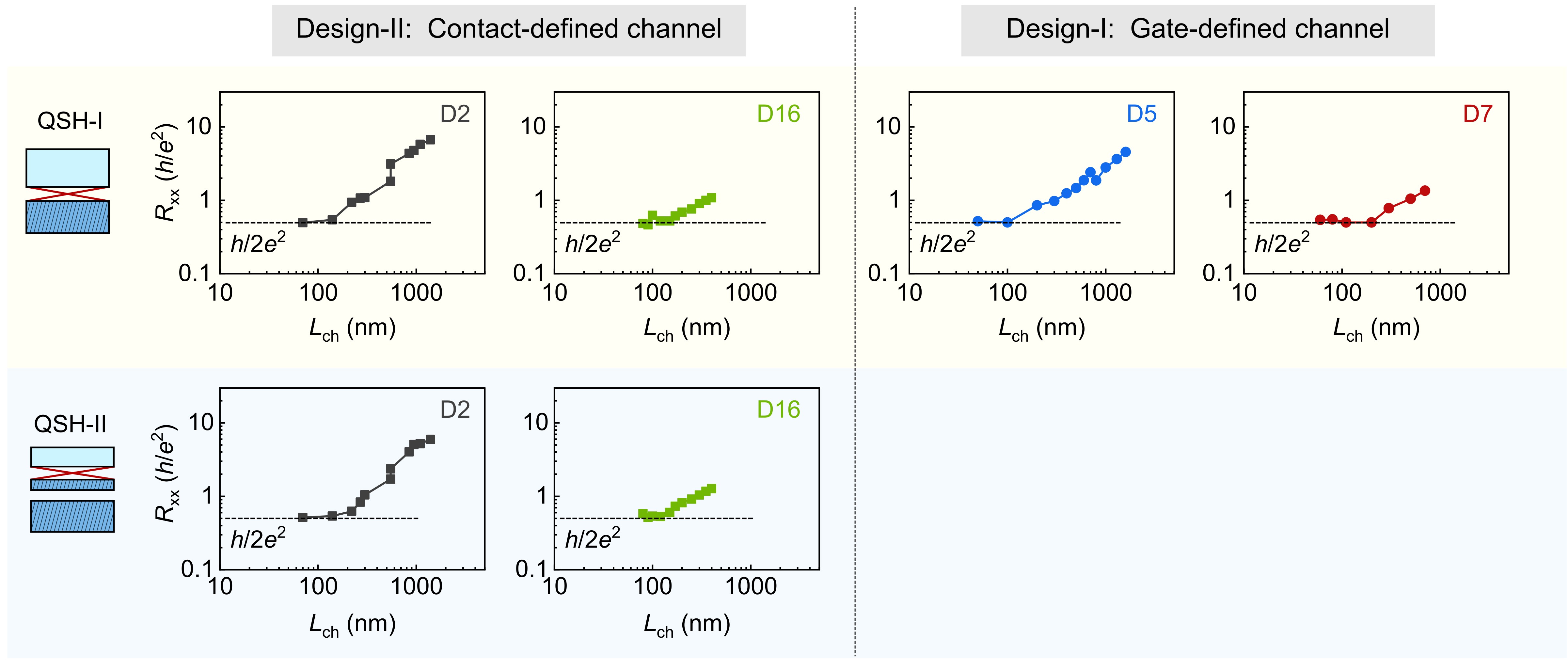}
\caption{{\figtitle{Channel length dependence of the resistance at the CNP gap (QSH-I) as well as the correlated gap (QSH-II).} Displayed here are the individual curves for devices D2, D5, D7, D16, as referenced in Fig.~\ref{Fig2}\textbf{g} and Fig.~\ref{Fig4}\textbf{e}.}}
\label{channel_length_seperated}
\end{figure*}

\begin{figure*}
\includegraphics[width=4.6in]{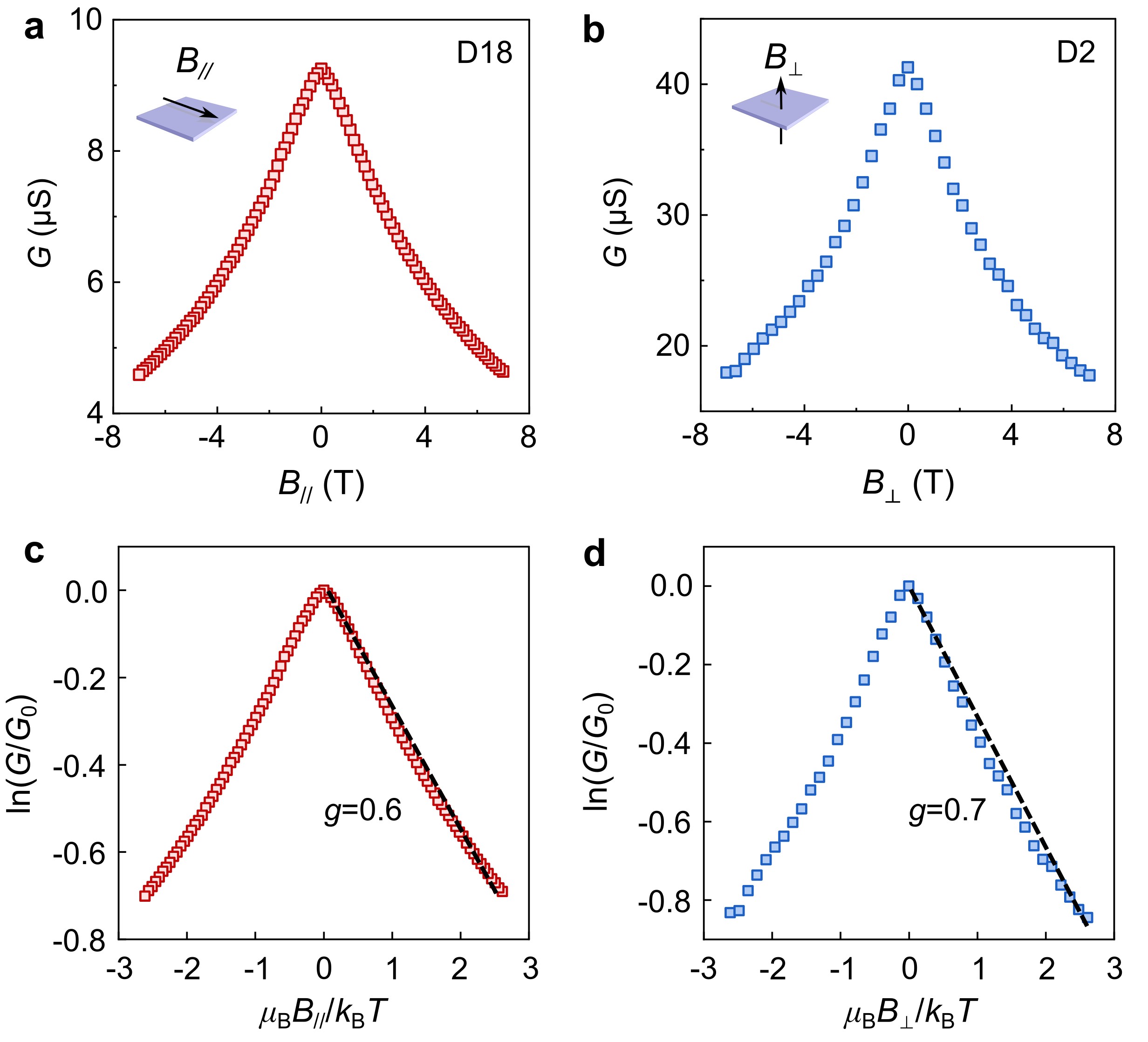}
\caption{{\figtitle{Conductance at CNP under in-plane and out-of-plane magnetic fields.} When a magnetic field is applied, a Zeeman gap is induced at the Dirac point of the edge states, influencing the edge conductance, which can be described by a thermal activation behavior: $G =G_0 e^{-g\mu_\textrm{B} |B|/2k_\textrm{B} T}$. Here, $G_0$ is the conductance at $B = 0$, $g$ is the effective g-factor, $\mu_\textrm{B}$ is the Bohr magneton, and $k_\textrm{B}$ is the Boltzmann constant. Panels (\textbf{a-b}) show the raw data $G$ versus $B$ for both in-plane and out-of-plane magnetic fields. Panels (\textbf{c-d}) plot the corresponding $\mathrm{ln}(G/G\mathrm{_0})$ versus $\mu\mathrm{_B}B/k\mathrm{_B}T$, from the slope of which the $g$-factor can be directly extracted.} The data was collected at $T$ = 1.7 K.}
\label{g_factor}
\end{figure*}

\begin{figure*}
\includegraphics[width=5in]{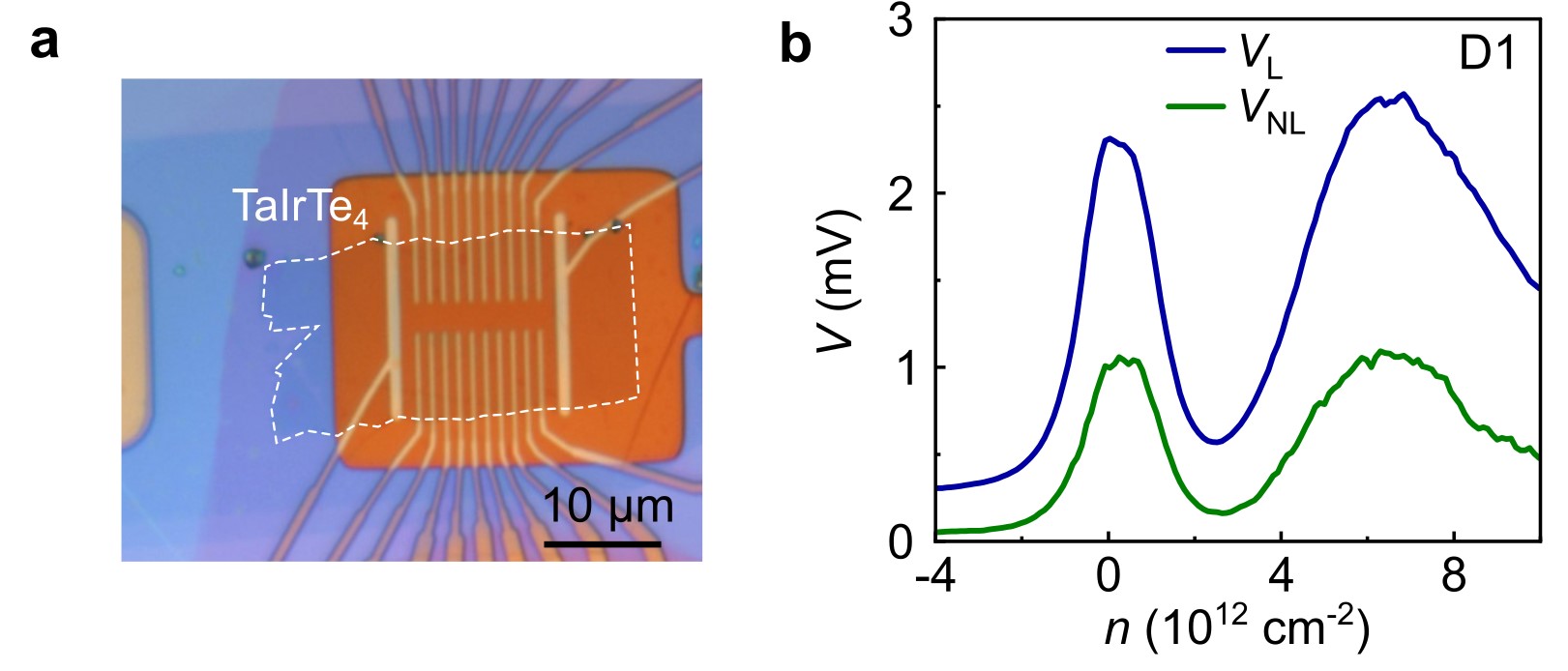}
\caption{{\figtitle{Enhanced nonlocal transport in both the CNP and second insulating gaps.} 
\textbf{a,} An optical image of device D1. Scale bar: 10 $\mu$m. \textbf{b,} The local ($V_\mathrm{L}$) and nonlocal ($V_\mathrm{NL}$) voltages measured with $I_\mathrm{xx}$ = 100 nA.  
}}\label{NL_eside}
\end{figure*}

\begin{figure*}
\includegraphics[width=5 in]{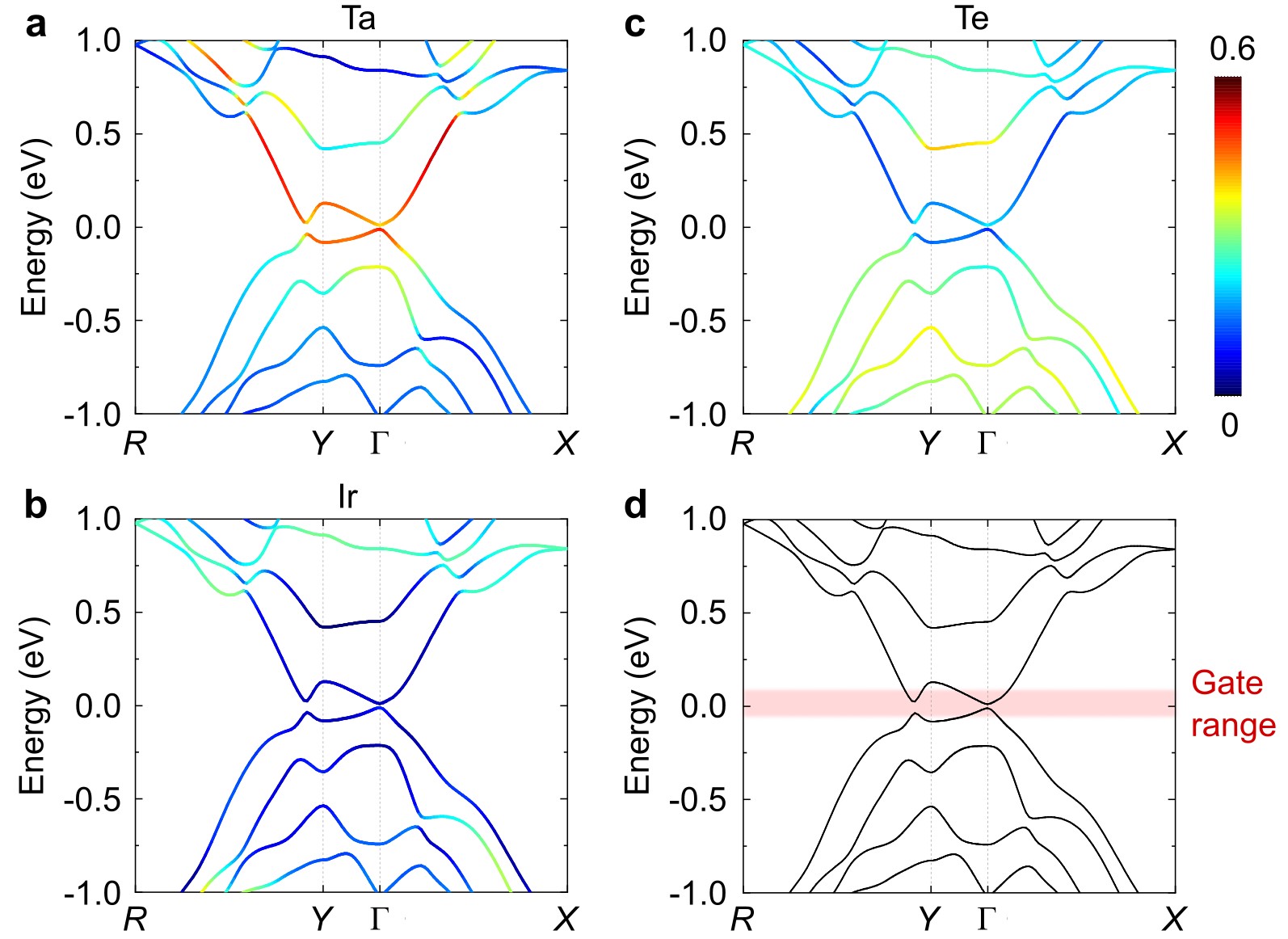}
\caption{{\textbf{Band structure and its orbital decomposition for monolayer TaIrTe$_4$.}
\textbf{a-c,} The orbital distribution of Ta \textbf{(a)}, Ir \textbf{(b)} and Te \textbf{(c)}, showing that the lowest energy bands originate primarily from Ta orbitals. \textbf{d,} The red-shaded region indicates the gate-tunable range within our experiment.}}\label{orbital}
\end{figure*}

\begin{figure*}
\includegraphics[width=6 in]{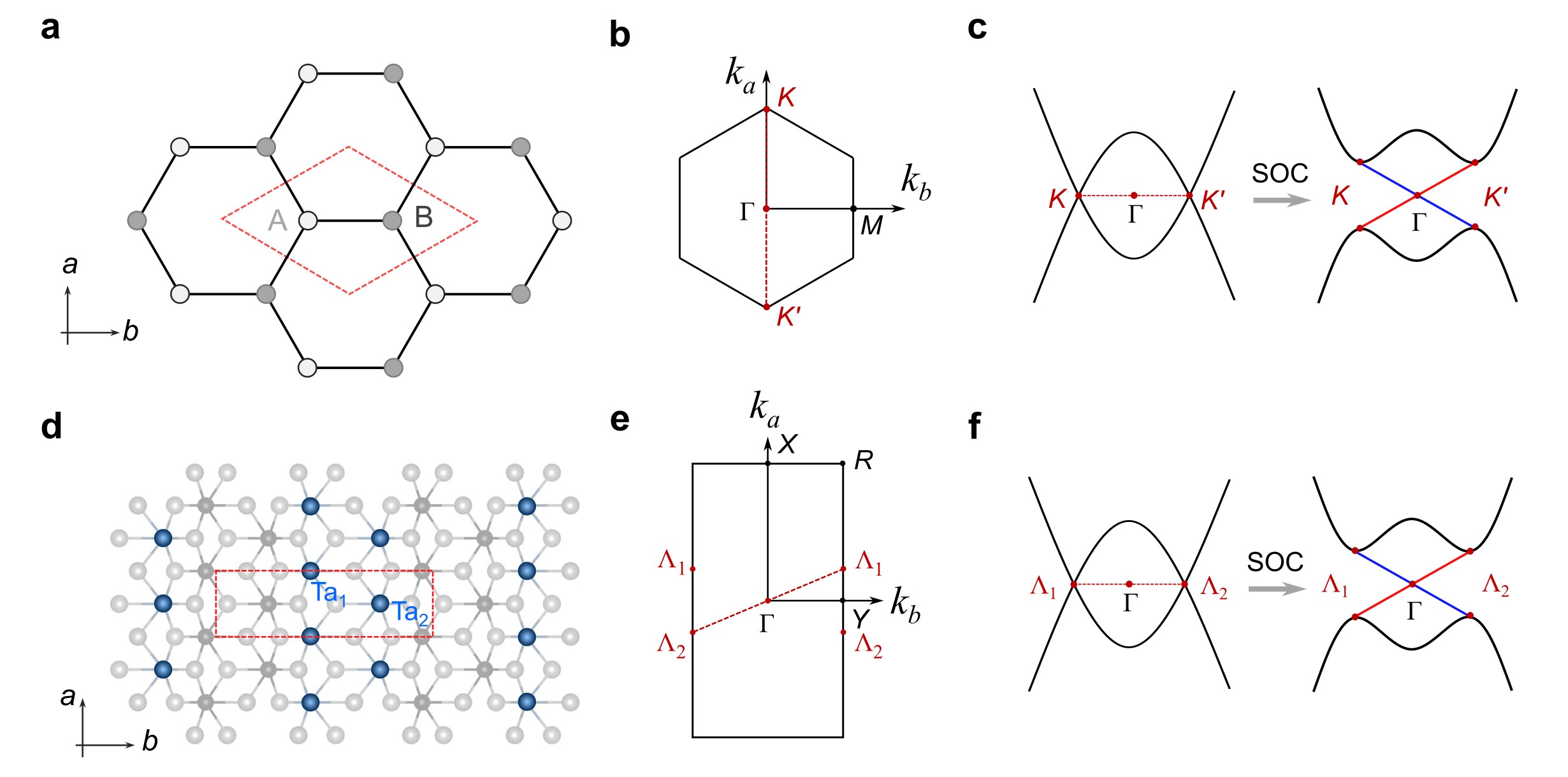}
\caption{{\figtitle{Formation of QSH bands in TaIrTe$_4$.}
\textbf{a,} Graphene lattice structure with two distinct atoms A and B per unit cell. 
\textbf{b,} Brillouin zone of graphene with $K$, $K'$ and $\Gamma$ points labeled. 
\textbf{c,} Existence of Dirac cones at $K$ and $K'$ points, where gap openings occur upon the introduction of spin-orbit coupling, leading to QSH edge states.
\textbf{d,} Lattice structure of monolayer TaIrTe$_4$ with two distinct Ta atoms (Ta$_1$ and Ta$_2$) within each unit cell. 
\textbf{e,} Brillouin zone of TaIrTe$_4$ with $\Lambda_1$, $\Lambda_2$, and $\Gamma$ points labeled.
\textbf{f,} Existence of Dirac cones at $\Lambda_1$ and $\Lambda_2$ points, where gap openings occur due to spin-orbit coupling, leading to QSH edge states.}}
\label{band_inversion}
\end{figure*}

\end{document}